\documentstyle[twocolumn,aps,epsf]{revtex}

\begin{document}

\twocolumn[
\hsize\textwidth\columnwidth\hsize\csname@twocolumnfalse\endcsname
\draft

\title{Entropy of vortex cores near the
superconductor-to-insulator transition in an underdoped cuprate}

\author{C. Capan, K. Behnia}
\address{Laboratoire de Physique Quantique(UPR5-CNRS), 
ESPCI, 10 Rue Vauquelin, F-75005 Paris, France}
\author{ J. Hinderer, A. G. M. Jansen}
\address{ Grenoble High Magnetic Field Laboratory(CNRS-MPI),  BP 166, F- 38042  Grenoble  , France }
\author{W. Lang }
\address{ Institut f\"ur Materialphysik, Universit\"at Wien, Boltzmanngasse 5
A-1090 Wien, Austria }
\author{C. Marcenat, C. Marin and  J. Flouquet}
\address{ DRFMC/SPSMS,  Commissariat \`a l'Energie Atomique, F-38042 
Grenoble, France }

\date{October 10, 2001}
\maketitle
	
\begin{abstract}
We present a study of Nernst effect in underdoped $La_{2-x}Sr_xCuO_4$ in 
magnetic fields as high as 28T. At high fields, a sizeable Nernst signal 
was found to persist in presence of a field-induced non-metallic resistivity. 
By simultaneously measuring resistivity and the Nernst coefficient, we extract the entropy of vortex cores in the vicinity of this 
field-induced superconductor-insulator transition. Moreover, the temperature 
dependence of the thermo-electric Hall angle provides strong constraints on the possible
origins of the finite Nernst signal above $T_c$, as recently discovered by
Xu {\it et al.}\cite{xu}. 
\end{abstract}

\pacs{}
]

An intriguing case of vicinity between superconducting and insulating ground
states occurs in the underdoped cuprates\cite{ando}. Various investigations
have shown that reducing the density of charge carriers\cite{semba} or
introducing disorder\cite{fukuzumi} or applying a magnetic field\cite{ando}
leads to the replacement of the superconductor with an insulator. The latter
route (i.e. the field-driven superconductor-to-insulator transition) raises
many unanswered questions, including the possible existence of vortices with
insulating cores. The structure of the vortex core in a doped Mott insulator
has been the subject of numerous theoretical studies\cite{arovas,han,franz}.
On the experimental side, recent neutron scattering experiments on La$_{2-x}$%
Sr$_{x}$CuO$_{4}$(LSCO) have revealed the existence of a dynamic magnetic
order associated with the vortex state which evolves towards a static order
in the underdoped regime\cite{lake}. Meanwhile, Scanning Tunnelling
Microscopy(STM) has proved to be a direct source of information on the
electronic spectrum inside the vortices, and have detected finite-energy
bound states in the vortex cores of optimally-doped cuprates\cite{pan,maggio}%
. Until now, however, experimental exploration of field-induced
superconductor-to-insulator transition has been limited to resistivity
measurements\cite{ando,karpinska}.

In this letter, we report on the evolution of the Nernst coefficient in
underdoped LSCO in the vicinity of the superconductor-to-insulator
transition. Nernst effect, the generation of a transverse electric field by
a longitudinal  thermal gradient in a magnetic field, has been an
instructive probe of vortex movement in the mixed state of high-T$_{c}$
superconductors in early 1990s\cite{ri}. Recently, Xu {\it et al.}\cite{xu}
reported the existence of a sizeable Nernst signal above T$_{c}$ over a
broad temperature range in underdoped LSCO. They argued that, due to the
cancellation of transverse currents generated by the thermal gradient and
the electric field in presence of a magnetic field, the quasiparticle
contribution to the Nernst signal should be negligible and interpreted their
finding as evidence for the existence of vortex-like excitations above T$%
_{c} $\cite{xu} in line with a scenario in which phase-coherent
superconductivity is destroyed above T$_{c}$ due to the weakness of phase
rigidity\cite{emery}. Our study, concentrated on the Nernst coefficient at
high magnetic field leads to several new findings. First, a large Nernst
signal was found to persist in presence of a field-induced non-metallic
behavior in resistivity. This observation provides new support for the
concept of vortices with insulating cores. Moreover, using both Nernst and
resistivity data, we calculate the entropy associated with the vortex cores
and compare it with the difference of entropy between the normal and
superconducting states as extrapolated from specific heat studies\cite{loram}%
. Finally, we present the first set of data on the thermoelectric Hall angle
and argue that its temperature-dependence (close to T$^{3}$) puts strong
constraints on the origin of the residual Nernst signal above T$_{c}$ first
discovered by Xu {\it et al.}\cite{xu}.

The preparation and characterisation of LSCO single crystals is described in
detail elsewhere\cite{marin}. Our set-up was designed in a way to measure
Seebeck and Nernst coefficients simultaneously as well as resistivity and
Hall effect. The temperature profile along the sample was monitored by two
miniature RuO$_{2}$ thermometers. Longitudinal and transverse DC voltages
produced by this heat current were measured by two EM N11 Nanovoltmeters.
The same contacts were used to measure electrical resistivity and Hall
coefficient. A superconducting magnet was used for experiments up to 12
teslas whereas a Bitter magnet at Grenoble High Magnetic Field Laboratory
was employed to access fields up to 28T.

Fig.1 shows the temperature-dependence of the Nernst signal and resistivity
in a La$_{1.92}$Sr$_{0.08}$CuO$_{4}$ single crystal for various magnetic
fields. In presence of a magnetic field of 12T, resistivity shows a broad
transition ending at T$\sim $5K. At the same field, we detect a large Nernst
signal which peaks at T$\sim $18K. Since the broad resistive transition is a
consequence of dissipation due to the vortex movement, a concomitant Nernst
signal due to the effect of a thermal force on the same vortices is
naturally expected. This is in agreement with what has been reported in the
case of optimally-doped cuprates\cite{ri}. However, the evolution of the
Nernst signal at higher fields is surprising. As seen in the upper panel of
Fig.1, a magnetic field of 26T is large enough to induce a slight
non-metallic behavior in resistivity in the 15K-40K temperature range.
However, at this field, the maximum in the Nernst signal occurs at almost
the same temperature, broadens, and presents a reduced but still large
magnitude. The coexistence of the peak in the Nernst signal with a
non-metallic resistivity is in sharp contrast for what has been reported for
all superconductors including optimally-doped cuprates\cite{ri}. It is the
main new finding of this letter. This result is confirmed on two other
single crystals at lower doping levels. Fig.2 shows the data on a La$_{1.94}$%
Sr$_{0.06}$CuO$_{4}$ single crystal. As seen in the figure, the sample shows
a very broad resistive transition at zero field. The application of a
magnetic field leads to the emergence of an insulating behavior, but barely
affects the peak in the Nernst signal which, nevertheless, presents a
reduced magnitude compared to the x=0.08 case. To reconcile the Nernst and
resistivity data, it is tempting to assume that at high field, the system is
populated by vortices which can move under the influence of a Lorentz force
and produce a non-metallic resistivity. This may arise in the context of an
insulating normal state, since the flux-flow resistivity, $\rho _{F}$ is a
fraction of the normal state resistivity $\rho _{N}$ (in the simplest case $%
\frac{\rho _{F}}{\rho _{N}}\propto \frac{H}{H_{c2}}$ \cite{bardeen}). Thus,
a non-metallic $\rho _{F}(T)$ may reflect the insulating behavior of $\rho
_{N}(T)$ with attenuation.

More insight on the fuzzy phase-boundary between the superconducting and the
normal states may be achieved by comparing the field-dependence of Nernst
effect and resistivity of the x=0.08 sample. As seen in the upper panel of
Fig.3, below T$_{c}$, Nernst coefficient is not a linear function of
magnetic field. It presents a maximum which becomes broader at lower
temperatures. Qualitatively, this behavior is understandable. The thermal
force on each vortex is proportional to the excess of entropy associated
with it. Since the latter would become zero at $H_{c2}$, the Nernst signal
is expected to decrease at a finite field in spite of the increase in the
number of vortices. As seen in the lower panel of Fig.3, the non-vanishing
Nernst signal is concomitant with a large magnetoresistance up to the
highest explored magnetic fields(28T). Extrapolating the Nernst data to
higher magnetic fields, one can estimate that the signal would vanish at H$%
\sim $60T which is close to the estimation of $H_{c2}$ deduced from
magnetoresistance saturation\cite{ando}. Fig.3 also displays the passage
between metallic and non-metallic behaviors at H$\sim $27T. It is worth
noting that the magnitude of resistivity at the boundary between metallicity
and non-metallicity(0.39 $m\Omega cm$) yields a resistance close to $\frac{h%
}{4e^{2}}$ per $CuO_{2}$ plane, reported as the critical threshold
resistance for superconductor-to-insulator transition in cuprates\cite{semba}%
.

Combining the Nernst and resistivity data, one can calculate the entropy
associated with these vortices\cite{ri}. When vortices move under the
influence of a thermal force: $f_{th}=-\frac{\partial T}{\partial x}S_{\phi
} $ ( $S_{\phi }$ is the transport entropy per unit length of an individual
vortex), they produce a transverse electric field according to the Josephson
equation, $E_{y}$=$B_{z}v_{x}$, where $v_{x}$ is the average vortex
velocity. This velocity is proportional to the force applied on a vortex $%
v_{x}=\eta f_{th}$. with $\eta $ being a viscosity coefficient. On the other
hand, flux movement in presence of a Lorentz force on individual vortices, $%
f_{L}=J_{x}\Phi _{0}$, is at the origin of the longitudinal electric field
produced by an electric current. $E_{x}$=$B_{z}v_{y}$. The same viscosity
coefficient relates this velocity to the Lorentz force $v_{y}=\eta f_{L}$.
Thus: 
\begin{equation}
\frac{J_{x}\Phi _{0}B_{z}}{E_{x}}=\frac{\frac{\partial T}{\partial x}S_{\phi
}B_{z}}{E_{y}}
\end{equation}

And defining resistivity as $\rho _{F}=\frac{E_{x}}{J_{x}}$ and the Nernst
coefficient as $N=\frac{E_{y}}{\frac{\partial T}{\partial x}}$, one obtains 
\cite{ri}:

\begin{equation}
S_{\phi }=\frac{N\Phi _{0}}{\rho _{F}}
\end{equation}

Now, the volume entropy at a given magnetic field is obtained by multiplying 
$S_{\phi }$ by $\frac{H}{\Phi _{0}}$, the density of the vortices at a given
field H. Hence, 
\begin{equation}
S_{m}=\frac{NH}{\rho _{F}V_{m}}
\end{equation}

is the excess entropy carried by vortices at a given field in molar units.
Here, V$_{m}$ is the molar volume (9.5 10$^{-29}$ m$^{3}$/mol for LSCO\cite
{harshman}). Fig.\ 4 displays the temperature dependence of $S_{m}$ obtained
in this way for H=12T and H=26T. In the picture sketched above, this plot
represents the difference between the entropy accumulated by the vortices
and the entropy of the background condensate.

As seen in the figure for both fields, $S_{m}$ shows a maximum and remains
finite well above T$_{c}$(=27K) in the ``normal'' state. This is a
consequence of the finite value of Nernst coefficient above T$_{c}$ and up
to the highest temperature explored in this study ($\sim $63K, see data in
the lower panel of Fig.1). This latter observation, first reported by Xu 
{\it et al.}\cite{xu} was interpreted by these authors as evidence for
vortex-like excitations in the pseudo-gap regime. So, the analysis sketched
above leads to the existence of a substantial $S_{m}$ persisting upto T*\cite
{xu} (the temperature below which the pseudo-gap opens up). Note that this
simple analysis remains valid for any exotic electronic excitation which
happens to be a reservoir of entropy (in order to move by a thermal
gradient) and either a topological defect in a phase-coherent environment or
a fluxoid (in order to produce an electric field by its movement). As
indicated by the absence of $\Phi _{0}$ in equation (3), it should not
necessarily be associated with a single flux quantum(i.e. a standard
Abrikosov vortex). Note, however, that while below T$_{c}$(H=0), the
resistivity is entirely generated by the vortex movement and there is no
ambiguity about the magnitude of $\rho _{F}$, this is not the case in the
normal state. Indeed, at this stage, in the absence of any solid evidence
for vortex dissipation in charge transport, the magnitude of $\rho _{F}$
above T$_{c}$ is a matter of speculation. On the other hand, our estimation
of $S_{m}$ in the superconducting state is straightforward and does not
suffer from the current uncertainty on the origin of the residual Nernst
signal in the pseudo-gap regime.

It is interesting to compare the temperature dependence and the magnitude of 
$S_{m}$ with the results of the extensive study of specific heat in cuprates
by Loram {\it et al}.\cite{loram}. Besides its finite value above T$_{c}$,
the overall temperature dependence of $S_{m}$(T) is reminiscent of the
difference between the entropy of the superconducting state and the
extrapolated entropy of the ``normal'' state obtained by specific heat
measurements. However, the magnitude of $S_{m}$ is almost an order of
magnitude smaller than the maximum in the difference in the specific heat
data for LSCO at this doping level\cite{loram}. The discrepancy is probably
due to the important differences between the nature of information obtained
by these two probes. First of all, our results are obtained in presence of a
strong magnetic field which is known to diminish and broaden the electronic
specific heat jump and consequently reduce the entropy difference between
the two states\cite{junod}. In the second place, the transport entropy of
(i.e. the entropy carried by) a vortex is yet to be theoretically clarified
in the context of d-wave superconductivity. To the first approximation, the
electronic excitation spectrum of the vortex core (defined as the region
within a coherence length of the center) reflects that of the normal state.
STM studies of the high-T$_{c}$ cuprates\cite{pan,maggio}, however, have
reported a remarkably reduced difference in the low-energy Density Of States
inside and far away from vortex cores compared to what has been observed in
a conventional superconductor\cite{hess}.

Finally, let us consider possible alternative origins of the observed Nernst
signal above T$_{c}$. Since vortex movement is not the only source of a
Nernst signal, it is useful to underline the restrictions that a
quasiparticle scenario should face in order to account for such a signal in
our context. For this purpose, we used our set-up to measure the ratio of
the longitudinal to transverse electric fields produced by a fixed thermal
current , $\overrightarrow{J_{q}}$, along the sample. This ratio directly
determines the field-induced rotation of the electric field produced by a
thermal current.  The cotangent of the thermoelectric Hall angle,  $\cot
\theta _{HTE}$, obtained in this way may be compared with the electric Hall
angle, $\cot \theta _{H}$. The lower panel of Fig.4 compares the evolution of
the two angles. As seen in the figure, the thermoelectric Hall angle
presents a T$^{3}$ behavior in the ``normal'' state and becomes even
stronger (and field-dependent) below T$_{c}$. This reflects the rapid
increase in the Nernst signal below T* and its non-linearity below T$_{c}$.
In the same temperature region, the measured electric Hall angle is almost
temperature-independent. [For LSCO at this doping level, the quadratic term
in $\cot \theta _{H}$ is very small at our temperture range\cite{balakirev}%
]. Now, $\overrightarrow{J_{q}}=\overline{\alpha }\overrightarrow{E}+%
\overline{\kappa }\overrightarrow{\nabla }T$ \ and , since there is no
charge current, $\overline{\sigma }\overrightarrow{E}=\frac{\overline{\alpha 
}}{T}\overrightarrow{\nabla }T$ which yields $\overrightarrow{J_{q}}=(%
\overline{\alpha }+\overline{\kappa }$ $\overline{\alpha }^{-1}\overline{%
\sigma }T)\overrightarrow{E}$ (see the inset of Fig. 1). Therefore the angle between  $%
\overrightarrow{J_{q}}$ and $\overrightarrow{E}$ reflects the  rotations
produced by  $\overline{\alpha }$, $\overline{\kappa }$ and $\overline{%
\sigma }$ which are the thermoelectric, thermal and electric conductivity
tensors. Explaining the rapid temperature-dependence of $\cot \theta _{HTE}$
seems to be a major challenge for the standard transport theory. Indeed, it
has  already been noted that in absence of an energy-dependence in the
scattering time, $\tau (\varepsilon )$, and even for a highly anisotropic
single Fermi surface, the two ratios $\frac{\alpha _{xx}}{\alpha _{xy}}$ and
$\frac{\sigma _{xx}}{\sigma _{xy}}$are expected to be equal\cite{clayhold}
which implies an identical rotation due to $\overline{\alpha }$ and 
$\overline{\sigma }$. More generally, in a Boltzmann picture\cite{ziman}: 
\begin{equation}
\overline{\alpha }=\frac{(\pi k_{B}T)^{2}}{3}\frac{\partial \overline{\sigma 
}}{\partial \epsilon }\mid _{\varepsilon =\varepsilon _{F}}
\end{equation}
which establishes an intimate relationship between the two angles even in
the case of a highly energy-dependent $\tau (\varepsilon ).$ Any alternative
scenario on the origin of the finite Nernst signal above T$_{c}$ implying
quasiparticles instead of vortices is expected to explain the contrasting
behavior of the two angles. We note that equation(4) is only valid when
charge and entropy are carried by the same electronic excitations which is
not the case in the charge-spin separation scenarios.

In summary, we studied Nernst effect at high magnetic fields in underdoped
LSCO and found that the Nernst signal persists in presence of a
magnetically-induced non-metallic behavior. We extracted the entropy carried
by vortices in the vicinity of this superconductor-to-insulator transition
and measured the temperature-dependence of the thermoelectric Hall angle.
This work was supported by the F\"orderung der wissenschaftlichen Forschung
of Austria, the Franco-Austrian Amadeus program and by the Fondation
Langlois.

\begin{figure}[tbp]
\epsfxsize=8.0cm
$$\epsffile{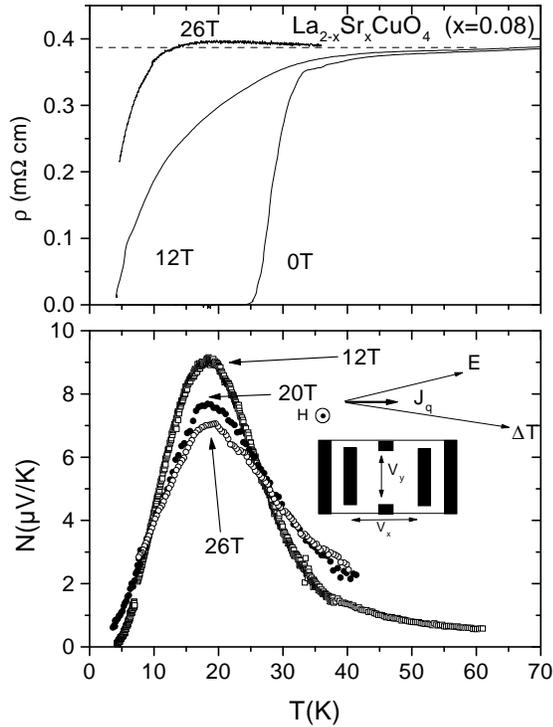}$$
\caption{Resistivity and Nernst effect as a function of temperature in an
underdoped LSCO crystal for different magnetic fields. The broken horizontal
line represents 0.39 $m\Omega cm$.  Inset shows the contact geometry on the sample
and the relevant vectors.}
\label{Fig. 1}
\end{figure}
\begin{figure}[tbp]
\epsfxsize=8.0cm
$$\epsffile{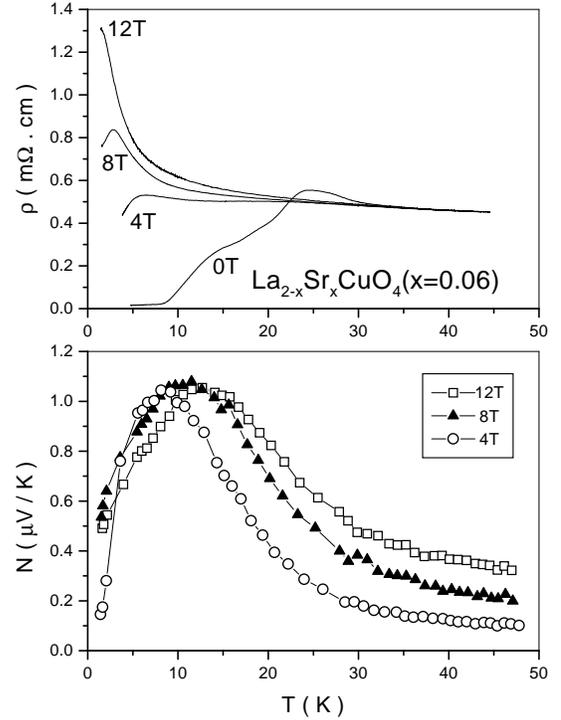}$$
\caption{Resistivity and Nernst effect as a function of temperature in
another underdoped LSCO crystal for different magnetic fields. Note the
coexistence of a Nernst signal and an apparently insulating behavior at
H=12T.}
\label{Fig. 2}
\end{figure}
\begin{figure}[tbp]
\epsfxsize=8.0cm
$$\epsffile{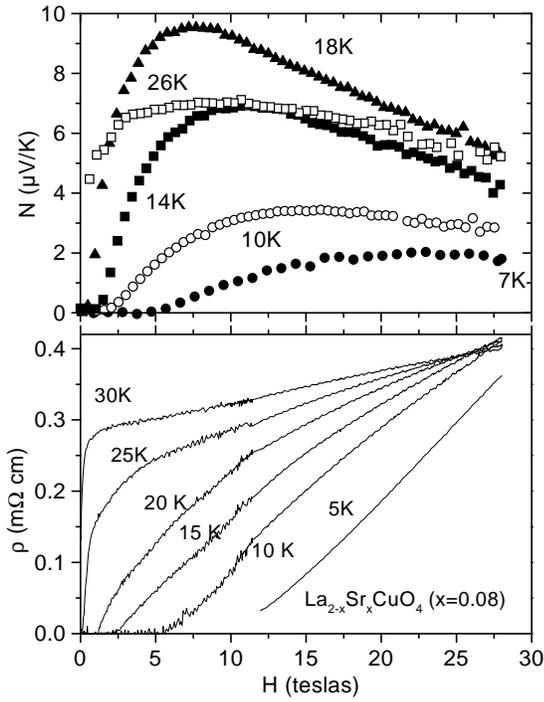}$$
\caption{Resistivity and Nernst effect as a function of field in an
underdoped LSCO crystal for different temperatures.}
\label{Fig. 3}
\end{figure}
\begin{figure}[tbp]
\epsfxsize=8.0cm
$$\epsffile{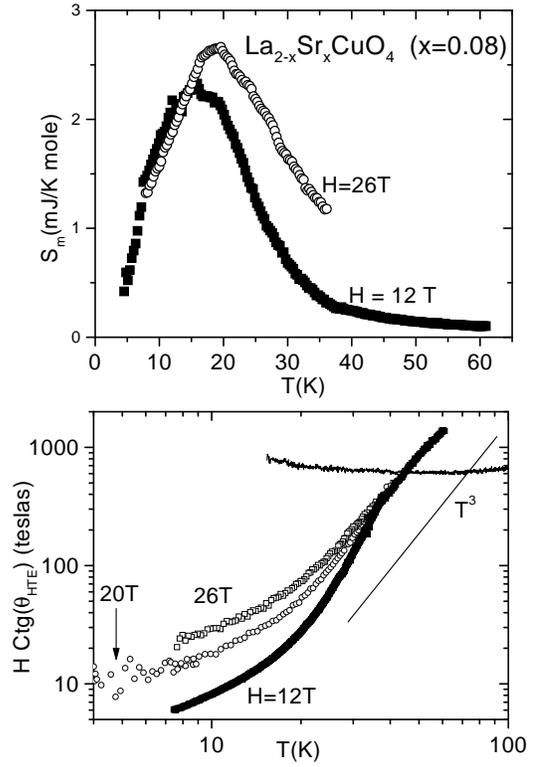}$$
\caption{Upper panel: Entropy carried by vortices as extracted from
resistivity and Nernst coefficient for the x=0.08 sample at two different
magnetic fields (See text).Lower Panel: The temperature dependence of the
normalised thermoelectric Hall angle at different magnetic fields. The solid
line represents the normalised electric Hall angle (measured at 12 T)for the
same sample. }
\label{Fig. 4}
\end{figure}

\end{document}